# IRS for Computer Character Sequences Filtration: a new software tool and algorithm to support the IRS at tokenization process

Ahmad Al Badawi
Department of Computers and Information Technology
Taif University
Taif, Saudi Arabia

Qasem Abu Al-Haija
Department of Electrical Engineering
King Faisal University
Alhasa, Saudi Arabia

*Abstract*—Tokenization is the task of chopping it up into pieces, called tokens, perhaps at the same time throwing away certain characters, such as punctuation. A token is an instance of token a sequence of characters in some particular document that are grouped together as a useful semantic unit for processing. New software tool and algorithm to support the IRS at tokenization process are presented. Our proposed tool will filter out the three computer character Sequences: IP-Addresses, Web URLs, Date, and Email Addresses. Our tool will use the pattern matching algorithms and filtration methods. After this process, the IRS can start a new tokenization process on the new retrieved text which will be free of these sequences.

## I. INTRODUCTION

People use search engines for instance to locate and buy goods, to choose a vacation destination, to select a medical treatment or to find background information on candidates of an election. It's necessary to build a searching system being able to support users expressing their searching by natural language queries is very important and opens the researching direction with many potential [7].

As a human nature, people prefer to search for their information using their natural language especially with the existence of huge amount of information these days. Thus, a good Information Retrieval System is on demand.

Information retrieval (IR) [1, 2] is finding material (usually documents) of an unstructured nature (usually text) that satisfies an information need from within large collections (usually stored on computers). An information retrieval system (IRS) [2] is a software program that stores and manages information on documents. This system assists users in finding the information they need. Designing an efficient IRS which tries to touch the optimal results in retrieving the relevant documents of information, that design will face a lot of parameters which are to be taken into account such as precision and recall [1]. There is always a trade of between precision and recall.

At the core of IRS immerges the tokenization process which is considered a primary part in IRSs. Tokenization [1] is the task of chopping character sequence up into pieces, called tokens. Sounds good; but it's not as simple as its definition, many times the tokenizer should not split on some locations of the document.

Computer technology [1] has introduced new types of character sequences that a tokenizer should probably tokenize as a single token, including email addresses (qalhaija@kfu.edu.sa), web URLs (http://www.kfu.edu.sa), numeric IP addresses (192.168.0.1), and Date (16/07/1982).

Several methods and approaches where proposed to provide these services, but researches showed that many measures should be addressed to ensure complete the retrieving process.

The problem addressed in this proposal will focus on designing an IRS for Computer Character Sequences Filtration. Computer technology has introduced new types of character sequences that a tokenizer should probably tokenize as a single token. As we see from figure 1, our proposed work will focus on filtering email addresses, web URLs, date, and numeric IP addresses.

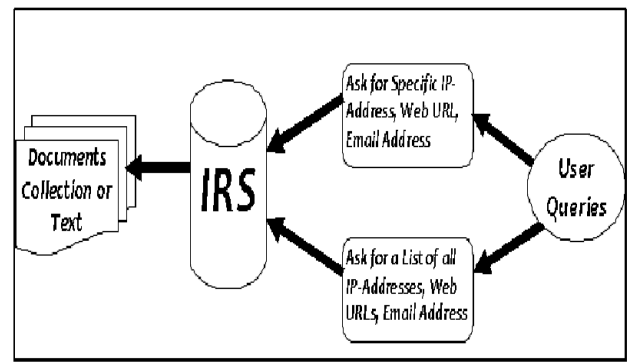

Fig. 1. Problem Statement Figure.

In this paper, we propose a new software tool and algorithm to support the IRS at tokenization process. Our proposed tool will filter out the three computer character Sequences: IP-Addresses, web URLs, and email addresses. Our tool uses the pattern matching algorithms [4, 5] and filtration methods [2, 3].

After this process, the IRS can start a new tokenization process on the new retrieved text which will be free of these sequences.

## II. Related Works

In the last years many classical solutions tried to address the tokenization process issues such as Computer Character Sequences Filtration. The most commonly used solution is the filtering and matching schemes [1].

Christopher D. Manning, Prabhakar Raghavan, Hinrich Schutze in [1] established the infrastructure for how to build IRSs by explaining all concepts and objects for several IRSs. They explained in details the tokenization process which sets at the core of the IR-Systems. They also discussed many of the challenges in the tokenization process such as the Computer Character Sequences (IP-Addresses, email addresses, date and web URLs), the use of the apostrophe for possession and contractions, hyphenation, foreign phrases, compound nouns and others.

Christos Faloutsos, Douglas Oard in [2] surveyed the major techniques for IRS. They provided an overview of some traditional IRSs (full text scanning, inversion, signature files and clustering) and discussed attempts to include semantic information (natural language processing, latent semantic indexing and neural networks.

Nicholas J. Belkin and W. Bruce Croft in [3] designed information filtering systems for unstructured or semi structured data, as opposed to database applications, which use very structured data. These systems also dealt primarily with textual information, but they may also entail images, voice, video or other data types that are parts of multimedia information systems.

Information filtering systems also involve a large amount of data and streams of incoming data, whether broadcast from a remote source or sent directly by other sources. Filtering is based on descriptions of individual or group information preferences, or profiles that typically represent long-term interests. Filtering also implies removal of data from an incoming stream rather than finding data in the stream; users see only the data that is extracted [1, 3].

Mary Elaine Califf, Raymond J. Mooney, in [4] presented an algorithm RAPIER, which uses pairs of sample documents and filled templates to induce pattern-match rules that directly extract fillers for the slots in the template. RAPIER is a bottom-up learning algorithm that incorporates techniques from several inductive logic programming systems. They have implemented the algorithm in a system that allows patterns to have constraints on the words, part-of-speech tags, and semantic classes present in the filler and the surrounding text. They presented encouraging experimental results on two domains.

Richard M. Karp, Michael O. Rabin, in [5] proposed randomized algorithms to solve the following string-matching problem and some of its generalizations. Given a string X of length n (the pattern) and a string Y (the text), find the first occurrence of X as a consecutive block within Y. The algorithms represent strings of length n by much shorter strings called fingerprints, and achieve their efficiency by manipulating fingerprints instead of longer strings. The algorithms require a constant number of storage locations, and essentially run in real time. They are conceptually simple and easy to implement. The method readily generalizes to higher-dimensional pattern-matching problems.

Sumalatha Ramachandran Sujaya Paulraj Sharon Joseph Vetriselvi and Ramaraj in [8] showed that there is no guarantee for information correctness and lots of conflicting information is retrieved by the search engines and the quality of provided information also varies from low quality to high quality. The filtering of trustworthiness is based on 5 factors - Provenance, Authority, Age, Popularity, and Related Links.

All the previous methods have strong filtration pattern-matching for the IR-Systems. Our proposed work is to design a new software tool and algorithm to support the IRS at tokenization process for computer character sequences filtration.

## III. Motivations and Methodology

The proposed research is motivated by many factors. First of all, the information retrieval which is a very important issue in real life applications as in banks, companies, hospitals, and at the personal level too. Second, previous studies showed that the tokenization process of IRS must leave some of the computer character sequences such as IP-Addresses, emails, web URLs, and date without any splitting operation; it should be treated as a single token. Our research will focus on the design and implementation of new software tool and algorithm to support the IRS at tokenization process. Another main reason to conduct this research is that IRSs are deemed world-wide hot research topic especially after the increasing demand on information and its applications.

The proposed methodology throughout this research consists of the following steps:

*1) The approach for proceeding in the proposed solution started by studying several Information Retrieval System (IRSs) [1, 2] concepts.*

*2) Understanding the important parameters in each IRS that can be helpful for solving the proposed problem.*

*3) Studying several information filtering systems [2, 3] which are designed for unstructured data.*

*4) Studying the matching algorithms [4, 5] trying to make some contribution to such algorithm. This will support the IR systems and will be helpful in solving our proposed problem.*

The proposed solutions were implemented and verified using Java programming language.

## IV. Simulation Environment

Our proposed work is to design a new software tool to support IRSs at the tokenization process to filter out some computer character sequences. Our proposed solution is programmed and implemented in Java programming language.

Java [6]; A simple, object-oriented, network-savvy, interpreted, robust, secure, architecture neutral, portable, high-performance, multithreaded, dynamic programming language. The Java programming language and environment is designed to solve a number of problems in modern programming practice. Java started as a part of a larger project to develop advanced software for consumer electronics. These devices are small, reliable, portable, distributed, and real-time embedded systems. When we started the project we intended to use C++, but encountered a number of problems. Initially they were just compiler technology problems, but as time passed more problems emerged that were best avoided by changing the language.

## V. TESTS AND RESULTS

The proposed work has shown marvelous results in terms of performance and low text processing using many advanced and intelligent techniques. Some of these techniques are regular expressions which is an advanced efficient technique for pattern matching. Regular expressions have been used to detect the special character sequences. What makes this work unique is the ability to extend and modify the application so that it can detect more character sequence patterns.

Another technique is the use of XML hierarchal structure in the configuration of the application. This has shown great results in terms of performance and use. The application can be extended to handle more and more special character sequences and can be integrated with many IRSs to boost their operations.

As mentioned previously, our principle is simulated in Java programming language. We have adopted two interfacing techniques to satisfy deferent requirements. The first is Command Line Interface (CLI) which is harder to use and interact but better in terms of performance. The other is the Graphical User Interface (GUI), the easier to use while worse when the performance has utmost priority. Anyway, we focus here on the GUI to show our work and results simply and clearly. Figure 2 shows the basic interface which will be the user's guide through the tokenization process.

Here is a brief description about each component, augmented with snapshots to make everything concrete:

- The "open" - shown in figure 3- button is used to load the ".txt" document that is intended to be tokenized. Upon clicking this button, open file dialog is popped up to the user to choose the desired document.

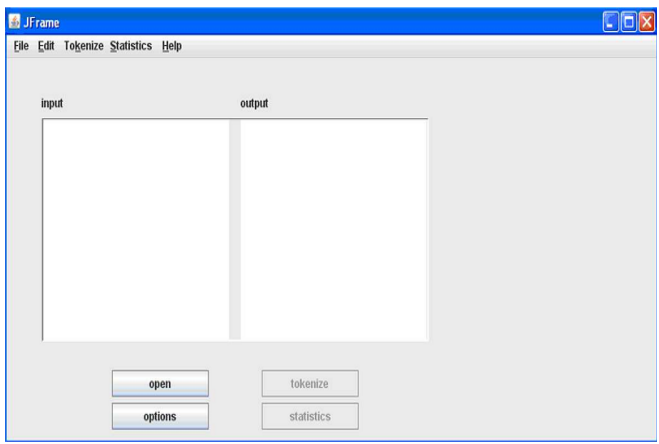

Fig. 2. The basic interface.

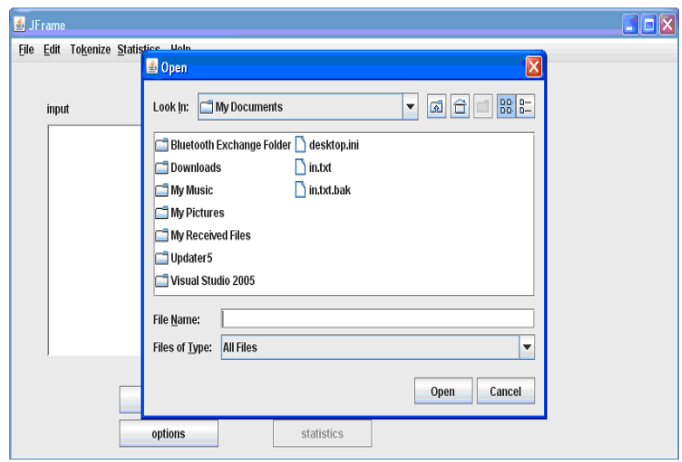

Fig. 3. The "open" button.

- Once the desired document is opened, see figure 4, the user is supposed now to choose the desired special character sequences to be recognized as one token. This can be accomplished by clicking the "options" button, or go to menu -> Tokenize -> options. As seen in figure 5 below, four checkboxes appears so that the user can choose the desired options.

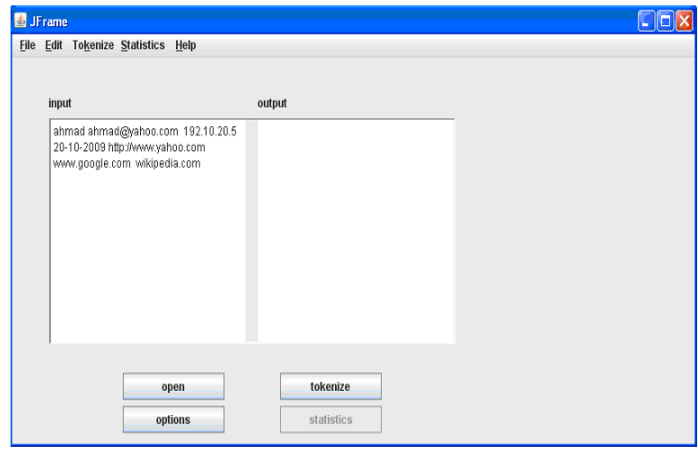

Fig. 4. Opened Document in the input pane.

- As we can see, there are four special character sequences: IP address, Email address, URL address, and Date. These special character sequences can be treated as one token, which could be removed (i.e. reduce the size of the index) or indexed as semantic unified terms which in turn increases the precession and the versatility of the IRS.

- Now, the user can initiate the tokenization process by clicking the tokenize button. The result of the tokenization will be saved in an output file (its path is configured) and will be shown to the user on the output text pane. Figure 5 below depicts this.

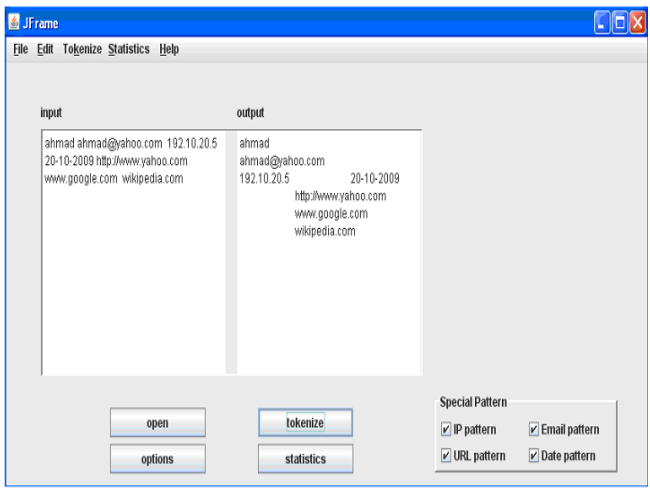

Fig. 5. Clicking the tokenize button

- Finally, the application provides one more feature that can be used in weight and score based IRSs. Upon clicking the statistics button, a piece of useful statistical results appears at the extreme right corner of the frame as shown below in figure 6.

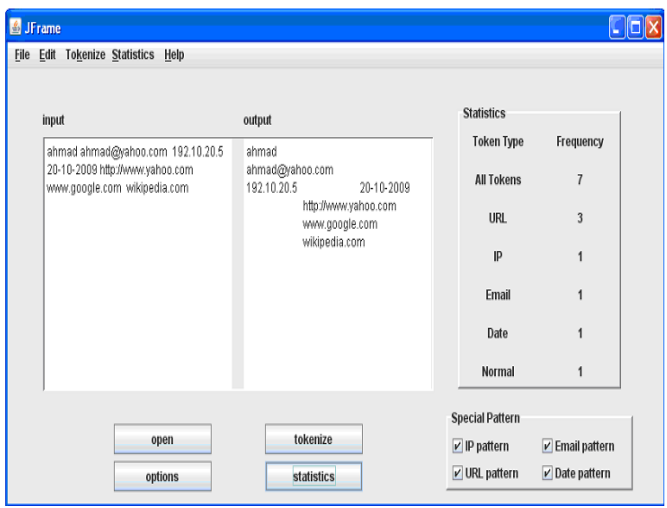

Fig. 6. Clicking on the statistics button.

## VI. CONCLUSIONS AND REMARKS

A new software tool and algorithm to support the IRS at tokenization process is implemented and proposed in this paper. Information retrieval Systems (IRS) which is meant of searching for information within documents and for metadata about documents. There are many applications of IR in the real life, for example Search engines like Google which sits at the throne of IRSs. At the heart of the IRS; is the tokenization process in which the text in the documents is split into small pieces called tokens. Sometimes there are some character sequences that must be taken as a single token without any splitting process, such as IP-Addresses, Email Address, date, and Web URLs.

This work can be extended by including other tokenization issues such as the use of the apostrophe for possession and contractions, hyphenation, foreign phrases, compound nouns and others. It can also be enhanced by implying some other issues such as stop words, normalization, stemming and lemmatization in information retrieval that can work in one coherent IR system.

Moreover, the ability to use heuristic algorithms such as Genetic algorithms that can make the IRS more efficient and can improve the system precision and recall. Finally, to involve retrieval tools that can be useful for bio-metrics and medical applications such as content-based image analysis [9].
ACKNOWLEDGEMENT

Authors would like to thank the deanship of scientific research at King Faisal University (KFU), Alhasa, Saudi Arabia for supporting this research.

AUTHORS PROFILE

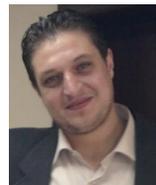

Eng. Qasem Abu Al-Haija' is a lecturer and Researcher at King Faisal University, College Of Engineering, Department of Electrical and Computer Engineering. He received his B.S. in Electrical and Computer Engineering from Jordanian Mu'tah University in February of 2005. Then he worked as a network engineer in a leading institute at KSA, and as a lecturer before he joined the graduate program at Jordan University of Science & Technology (JUST) in September 2007. Eng. Qasem received his M.S. degree in Computer engineering from Jordan University of Science & Technology in December 2009. Eng. Qasem research interests include Cryptography and Security, Computer Arithmetic and Finite Fields, Hardware implementations for cryptography, Wireless Sensor Networks, FPGA design, Elliptic Curve Cryptography, computer architecture, digital arithmetic algorithms.

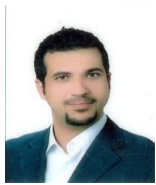

Eng. Ahmad Al Badawi is a lecturer and Researcher at Taif University, College Of Computers and Information Technology, Department of Computer Engineering. He received his B.S. in Electrical and Computer Engineering from the Faculty of Engineering Technology, Al-Balqa Applied University in June 2007. Then he worked as Sr. Software Developer at Globitel, a leading Converged Telecommunication Solutions provider, in Jordan before he joined the graduate program at Jordan University of Science & Technology (JUST) in September 2007. Eng. Ahmad received his M.S. degree in Computer engineering from Jordan University of Science & Technology in March 2010. Eng. Ahmad research interests include Particle Swarm Optimization, Multiprocessor Scheduling, Parallel Processing, Information Security and Cryptography, and Wireless Sensor Networks.